\documentclass[preprintnumbers,amsmath,amssymb,showpacs,twocolumn]{revtex4}

\usepackage{graphicx}
\usepackage{dcolumn}
\usepackage{bm}

\begin{document}

\title{Optimal Controlled Teleportation}

\author{Ting Gao$^{1,2}$, Feng-Li Yan$^{2,3}$, and You-Cheng Li$^3$}

\affiliation {$^1$ College of Mathematics and Information Science, Hebei Normal University, Shijiazhuang 050016, China\\
$^2$ Max-Planck-Institut f\"{u}r Quantenoptik, Hans-Kopfermann-Str. 1, D-85748 Garching, Germany \\
$^3$ College of Physics and Information Engineering, Hebei Normal University, Shijiazhuang 050016, China}

\date{\today}

\begin{abstract}
 We give the
analytic expressions of  maximal probabilities of successfully controlled teleportating an unknown qubit via
every kind of tripartite states. Besides, another kind of localizable entanglement  is also determined.
Furthermore, we give the sufficient and necessary condition that a three-qubit state can be collapsed to an EPR
pair by a measurement on one qubit, and characterize the three-qubit states that can be used as quantum channel
for controlled teleporting a qubit of unknown information with unit probability  and with unit fidelity.
\end{abstract}


\maketitle

Quantum teleportation is commonly considered as one of the most striking progress of quantum information theory.
In the seminal work of Bennett \emph{et al.} \cite{BBCJPW}, they showed that an arbitrary unknown state of a
qubit could be teleported from a sender Alice to a spatially distant receiver Bob with the aid of long-range
Einstein-Podolsky-Rosen (EPR) correlations and the transmission of two bits of classical information. Since
then, quantum teleportation has been developed by many authors due to its important applications in quantum
communication
 and quantum computation. At present, teleportation has been generalized to various cases
  \cite{JBS, VaidmanPRA1994, BKprl1998,  BDEFMS,  KB,YCHpra2004, PatiPRA2000, PAjob, DLLZWpra2005}.
  On the other hand, in past several years quantum teleportation has been also experimentally
 demonstrated by several groups \cite{BPMEWZ,NKL}.

  The controlled quantum
 teleportation scheme  was  presented by Karlsson and Bourennane \cite{KB}, with very similar ideas also in the quantum secret sharing
  paper of Hillery et al. \cite{HBB}. In \cite{KB,HBB} the entanglement property of the Greenberger-Horne-Zeilinger (GHZ)
state  is utilized for teleporting a qubit of unknown information.  According to the scheme, a third side is
included, so that the quantum channel is supervised by this additional side. An unknown state can be perfectly
transported from one place to another place via previously shared quantum resourse---GHZ state---by means of
local operations and classical communications (LOCC) under the permission of the third party. The signal state
can not be transmitted unless all three sides agree to cooperate. The controlled quantum teleportation is useful
in the context of quantum information such as networked quantum information processing and cryptographic
conferencing \cite{ BHMpra1996, Townsend, BVKpra2003}, and controlled quantum secure direct communication  and
has other interesting applications, such as in opening account on the agreement of managers in a network.
Recently, a number of works on controlled quantum teleportation have also been proposed \cite{YCHpra2004,
PatiPRA2000, DLLZWpra2005}, where they restrict themselves to the special quantum channels, such as GHZ state or
W state. If a nonmaximally entangled state is taken as quantum channel, then one can not teleport a qubit with
unit probability  and unit fidelity. However, it is possible to teleport a qubit with a probability $p<1$, which
is called probabilistic quantum teleportation \cite{PAjob}.

The entanglement property lies at the very heart of quantum information theory. The reason is that entanglement is the physical resource to
perform some of the most important quantum information tasks, such as quantum teleportation, quantum computation etc. In \cite{VPCirac-prl(04)},
Verstraete, Popp, and Cirac introduced a new concept which they called localizable entanglement (LE). This quantity not only has a very well
defined physical meaning that treats entanglement as a truly physical resource, but also establishes a very close connection between
entanglement and correlation functions. The LE of two particles is the maximal amount of entanglement that can be localized in these two
particles, on average, by doing local measurement on the rest of the particles. The determination of the LE is a formidable task since it
involves optimization over all possible local measurement strategies, and thus can not be determined in general. However,  Verstaete, Popp, and
Cirac  gave  tight upper bound and lower pound.

In this paper, we investigate  the general case of controlled quantum teleportation --- i.e. controlled
teleporting a qubit of unknown information from a sender to a remote receiver via the control of a third agent
by the use of a general three-qubit state --- and its maximal successful probability, which is a kind of  LE,
but different from that in \cite{VPCirac-prl(04)}.  We give the analytic expression  of the maximal successful
probability and the exact value of LE in \cite{VPCirac-prl(04)} for tripartite. Moreover, the sufficient and
necessary condition that a general three-qubit states can collapsed to an EPR pair with certain probability by
means of measurement on one qubit  are given.  In addition, we show in detail that for any given three-qubit
state, how to choose measurement basis to achieve maximal successful probability of controlled teleportation.
More surprising is the fact that there exist states that can not be converted to GHZ states under LOCC and can
be used for perfect controlled teleportation
--- controlled teleporting a qubit with unit fidelity and unit probability. In deed, we show that any tripartite entangled state can be used for
perfect controlled teleportation if and only if it is LOCC equivalent to the following state
\begin{eqnarray}\label{extendedGHZ}
 a_0|000\rangle+a_1|100\rangle+ \frac{1}{\sqrt{2}}|111\rangle,
\end{eqnarray}
 where  $a_0\geq 0$, $a_1\geq 0$, and $a_0^2+a_1^2=\frac{1}{2}$. Here the first qubit is a controlled one.

Ac\'{i}n  et al \cite{AACJLT}  proved that for every pure state of a composite system, 123, there exist
orthonormal states $|0\rangle_1$, $|1\rangle_1$ for system 1, orthonormal states $|0\rangle_2$, $|1\rangle_2$
for system 2, and orthonormal states $|0\rangle_3$, $|1\rangle_3$ for system 3  such that
\begin{equation}\label{quantumchannel}
\begin{array}{ll}
|\Psi\rangle_{123}= & a_0|000\rangle_{123}+a_1e^{\texttt{i}\mu}|100\rangle_{123}+a_2|101\rangle_{123} \\
  & +a_3|110\rangle_{123}+a_4|111\rangle_{123}, \\
 &  a_i\geq 0, ~ 0\leq
  \mu\leq \pi, ~ \Sigma_{i=0}^{4} a_i^2=1.
\end{array}
\end{equation}

Suppose that Alice is to deliver an unknown state $|\psi\rangle_4=\alpha|0\rangle_4+\beta|1\rangle_4$ ( $|\alpha|^2+|\beta|^2=1$) to a distant
receiver Bob supervised by the controller Charlie  via a quantum channel of a normalized general pure three-qubit
 state in (\ref{quantumchannel}),
 where particle 1 belongs to Charlie, particle 2 is in Alice's
 side, while Bob has particle 3.  Note that $a_0\neq 0$. Since if $a_0=0$, then $|\Psi\rangle_{123}$ is a tensor product state of a pure state of
 particle 1 and a pure state of particles 2 and 3, but not a true tripartite entangled state.

 After  getting the
approval of Charlie, Alice and Bob begin their teleportation under the control of Charlie.

The controller Charlie  measures his particle in the measurement basis
\begin{equation}\label{measurebasis}\begin{array}{l}
|x\rangle=\cos(\theta/2)|0\rangle+e^{\texttt{i}\varphi}\sin(\theta/2)|1\rangle,\\
|x\rangle^\perp=\sin(\theta/2)|0\rangle-e^{\texttt{i}\varphi}\cos(\theta/2)|1\rangle,
\end{array}\end{equation}
and broadcasts his measurement result to  Alice and Bob. Here $ \theta\in[0,\pi], \varphi\in[0,2\pi]$.

The tripartite state $|\Psi\rangle_{123}$ can be reexpressed as
\begin{equation}\label{quantumchanneldecomposition}
\begin{array}{lll}
|\Psi\rangle_{123} &=& \sqrt{p_1}|x\rangle_1|\Phi_1\rangle_{23}+\sqrt{p_2}|x\rangle_1^\perp|\Phi_2\rangle_{23}.
\end{array}
\end{equation}
Here {\small
\begin{eqnarray*}
 p_1  &=& \sin^2(\theta/2)+a_0^2\cos\theta+a_0a_1\cos(\mu-\varphi)\sin\theta, \label{p1}\\
  p_2 &=& \cos^2(\theta/2)-a_0^2\cos\theta-a_0a_1\cos(\mu-\varphi)\sin\theta, \label{p2}\\
 |\Phi_1\rangle_{23} &=& p_1^{-\frac{1}{2}}\{[a_0\cos(\theta/2)+a_1e^{\texttt{i}(\mu-\varphi)}\sin(\theta/2)]|00\rangle_{23} \nonumber\\
&&   +e^{-\texttt{i}\varphi}\sin(\theta/2)[a_2|01\rangle_{23}+a_3|10\rangle_{23}
  +a_4|11\rangle_{23}]\}, \\
  |\Phi_2\rangle_{23} &=& p_2^{-\frac{1}{2}}\{[a_0\sin(\theta/2)-a_1e^{\texttt{i}(\mu-\varphi)}\cos(\theta/2)]|00\rangle_{23} \nonumber\\
&& -e^{-\texttt{i}\varphi}\cos(\theta/2)[a_2|01\rangle_{23}+a_3|10\rangle_{23}  +a_4|11\rangle_{23}]\}.
\end{eqnarray*}}

After Charlie's measurement, the quantum channel is collapsed to $|\Phi_1\rangle_{23}$, and  $|\Phi_2\rangle_{23}$ with probability $p_1$, and
$p_2$, respectively.

 By Schmidt decomposition, there is
\begin{eqnarray}\label{schmidtdecomposition}
|\Phi_1\rangle_{23}&=& \sqrt{\lambda_{10}}|0_2'0_3'\rangle+\sqrt{\lambda_{11}}|1'_21'_3\rangle,\\
|\Phi_2\rangle_{23}&=& \sqrt{\lambda_{20}}|\bar{0}_2\bar{0}_3\rangle+\sqrt{\lambda_{21}}|\bar{1}_2\bar{1}_3\rangle,
\label{schmidtdecomposition1}
\end{eqnarray}
where $\{0_2',1_2'\}$ and  $\{\bar{0}_2,\bar{1}_2\}$ ( $\{0_3',1_3'\}$, and  $\{\bar{0}_3,\bar{1}_3\}$ ) are orthonormal bases of system 2
(system 3), and
\begin{equation}\label{}
\begin{array}{c}
  \lambda_{10}=\frac{1-\sqrt{1-C_1^2}}{2}, ~  \lambda_{11}= \frac{1+\sqrt{1-C_1^2}}{2}, \\
  \lambda_{20}= \frac{1-\sqrt{1-C_2^2}}{2}, ~  \lambda_{21} = \frac{1+\sqrt{1-C_2^2}}{2},\\
  C_1 = \frac{|a_0a_4e^{-\texttt{i}\varphi}\sin\theta+2(a_1a_4e^{\texttt{i}\mu}
-a_2a_3)e^{-2i\varphi}\sin^2\frac{\theta}{2}|}{p_1}, \label{c1} \\
  C_2= \frac{|a_0a_4e^{-\texttt{i}\varphi}\sin\theta-2(a_1a_4e^{\texttt{i}\mu}
-a_2a_3)e^{-2i\varphi}\cos^2\frac{\theta}{2}|}{p_2}.
\end{array}
\end{equation}

Then, Alice makes a Bell measurement on her particles 2 and 4, and conveys  her measurement outcome to Bob by transmitting two classical bits of
information  over a classical communication channel.

 In order to achieve teleportation, Bob needs to introduce an auxiliary particle $b$ with
the initial state $|0\rangle_b$ and performs a collective unitary on the state of particles 3 and $b$. Then the measurement on his auxiliary
particle $b$ follows. If his measurement result is $|0\rangle_b$, Bob can fix up the state of his particle 3, recovering $|\psi\rangle$, by
applying an appropriate local unitary operation.   The achievable successful probability of teleporting the unknown state  via
$|\Phi_1\rangle_{23}$ ($|\Phi_2\rangle_{23}$)  is $2\lambda_{10}$ ($2\lambda_{20}$).

Probability $p$ of successfully controlled teleporting an unknown qubit  using the state in (\ref{quantumchannel}) is
\begin{equation}\label{probability}
   p=  2p_1\lambda_{10}+2p_2\lambda_{20}= 1-R(\theta,\varphi).
\end{equation}
Here
\begin{equation}\label{}
\begin{array}{c}
R= R(\theta,\varphi)=\sqrt{P(\theta,\varphi)} + \sqrt{Q(\theta,\varphi)}, \\
  P(\theta, \varphi)= p_1^2(1-C_1^2), ~ Q(\theta,\varphi) = p_2^2(1-C_2^2).
\end{array}
\end{equation}

It is known that if there exists $(\theta_0, \varphi_0)$  such that  $P(\theta_0,\varphi_0)=0$ or
$Q(\theta_0,\varphi_0)=0$, then Alice and Bob share an EPR pair with some finite probability by Charlie's
measurement in the basis (\ref{measurebasis}) with $(\theta, \varphi)=(\theta_0, \varphi_0)$.
 Furthermore, if there exists $(\theta_0, \varphi_0)$  such that  $P(\theta_0,\varphi_0)=Q(\theta_0,\varphi_0)=0$, then an EPR pair occurs with
certainty after Charlie's measurement in the basis (\ref{measurebasis}) with $(\theta, \varphi)=(\theta_0, \varphi_0)$. The tripartite state
(\ref{quantumchannel}) with the property that
 both $P(\theta,\varphi)$ and $Q(\theta,\varphi)$ are equal to zero at the same point $(\theta,\varphi)$ can be used for perfect
teleportation.

We first investigate the condition of $P(\theta,\varphi)=0$.

$P(\theta,\varphi)=0$ $\Longleftrightarrow$ $|\Phi_1\rangle_{23}$ is a  Bell state $\Longleftrightarrow$ the
concurrence $C_1$ of $|\Phi_1\rangle_{23}$ is 1.

Since $P(0,\varphi)=a_0^2\neq 0$, so we suppose $\theta\in(0,\pi]$.   Let
$|\phi_1\rangle_{23}=\frac{ae^{\texttt{i}\alpha}|00\rangle_{23}
   +a_2|01\rangle_{23}+a_3|10\rangle_{23}+a_4|11\rangle_{23}}{\sqrt{a^2+a_2^2+a_3^2+a_4^2}}.$
Here $ ae^{\texttt{i}\alpha}\equiv t a_0e^{\texttt{i}\varphi}+a_1e^{\texttt{i}\mu},$ and $t=\cot\frac{\theta}{2},$ while $a$ is the absolute of
complex number $ ae^{\texttt{i}\alpha}$,  $\alpha$ is the argument of $ae^{\texttt{i}\alpha}$. Note that the concurrence $C_1$ of
$|\Phi_1\rangle_{23}$ is equal to the concurrence $C(|\phi_1\rangle_{23})$ of $|\phi_1\rangle_{23}$, and $C(|\phi_1\rangle_{23})=1$ iff
$[C(|\phi_1\rangle_{23})]^2=1$. $[C(|\phi_1\rangle_{23})]^2=1$ means that
\begin{equation}\label{square-of-concurrence =1}
\begin{array}{ll}
 & (a^2-a_4^2)^2+2(aa_2-a_3a_4)^2+2(aa_3-a_2a_4)^2\\
& +8aa_2a_3a_4(1+\cos\alpha)+(a_2^2-a_3^2)^2=0.
\end{array}
\end{equation}
From Eq.(\ref{square-of-concurrence =1}), we  see that $C_1=1$ if and only if $a_2=a_3$, $a=a_4$, and
$\alpha=\pi$. Using these equalities, we obtain the following results: If $a_1=0$, then $\varphi=\pi$ and
$t=\frac{a_4}{a_0}$. If $\mu=0$, then $\varphi=\pi$ and $t=\frac{a_1+a_4}{a_0}$. If $\mu=\pi$, then either
$\varphi=0$ and $t=\frac{a_1-a_4}{a_0}$ in case of $a_1>a_4$ or $\varphi=\pi$ and $t=\frac{a_4-a_1}{a_0}$ in
case of $a_1<a_4$. If $a_1\sin\mu\neq 0$, then $\cot\frac{\theta}{2}
=\frac{\sqrt{a_1^2+2a_1a_4\cos\mu+a_4^2}}{a_0},$ $ \cot\varphi =\cot\mu+\frac{a_4}{a_1\sin\mu},
\varphi\in(\pi,2\pi).$

 Similarly, we derive that
$Q(\theta,\varphi)=0$ if and only if the coefficients of the tripartite state in (\ref{quantumchannel}) satisfy
$a_2=a_3$.

Thus, we find that  a three-qubit state in (\ref{quantumchannel}) can be collapsed to an EPR pair with certain probability  by a  measurement on
the first qubit  if and only if  $a_2=a_3$.

Next we characterize the states such that both $P(\theta,\varphi)$ and $Q(\theta,\varphi)$ are equal to 0 at the
same point $(\theta,\varphi)$.

By above discussion, we need only to find the condition such that $Q(\theta_0,\varphi_0)=0$ for each case with
$P(\theta_0,\varphi_0)=0$. Note that $a_0\neq 0$. We see that
$Q(\theta_0,\varphi_0)=(1-2a_3^2-2a_4^2)^2+4a_3^2(a_1-a_4)^2=0$
 if and only if $a_2=a_3=0$ and $a_4=\frac{1}{\sqrt{2}}$  in case of  $a_1\sin\mu=0$ and $a_2=a_3$.
  Similarly,  we have that  when $a_2=a_3$, and $a_1\sin\mu\neq 0$,
   $Q(\theta_0,\varphi_0)=(1-2a_3^2-2a_4^2)^2+4a_3^2(a_1^2+2a_1a_4\cos\mu+a_4^2)=0$
 if and only if  $a_2=a_3=0$ and $a_4=\frac{1}{\sqrt{2}}$.

Therefore, three-qubit state in the generalized Schmidt decomposition (\ref{quantumchannel}) can be used for perfect  teleportation  if and only
if it is the state (\ref{extendedGHZ}).

Now we investigate how to  achieve the maximum of probability of successfully controlled teleporting an unknown
qubit state via an arbitrary partially entangled quantum channel (\ref{quantumchannel}).

Obviously, the maximum of (\ref{probability}) is
\begin{equation}\label{max}
 \begin{array}{c}
p_{\max}=  \max\{p\} = 1-\min\{R(\theta,\varphi)\}=1-R_{\min}.
 \end{array}
\end{equation}

In order to reach the maximal probability of exact controlled teleportation, the supervisor Charlie needs only
to choose optimal measurement basis, i.e. he selects $\theta_{\bar{0}}$ and $\varphi_{\bar{0}}$ such that
$R_{\min}=R(\theta_{\bar{0}},\varphi_{\bar{0}})$

Note that  the minimum of $R(\theta,\varphi)$ should occur at the points such that $P(\theta,\varphi)=0$, $Q(\theta,\varphi)=0$, and
\begin{eqnarray} \label{differencial-theta-and-varphi}
 R'_\theta(\theta,\varphi)= \frac{\partial  R}{\partial \theta}=0, ~
  R'_\varphi(\theta,\varphi)=  \frac{\partial R}{\partial \varphi}= 0.
\end{eqnarray}
Combining these two equations gives {\small
\begin{eqnarray}
 \frac{\partial P}{\partial\theta}\frac{\partial Q}{\partial\varphi}
   -\frac{\partial Q}{\partial\theta}\frac{\partial P}{\partial\varphi} &=& 0, \label{diff-theta-varphi}\\
   P\left(\frac{\partial Q}{\partial\varphi}\right)^2
   -Q\left(\frac{\partial P}{\partial\varphi}\right)^2 &=& 0, \label{diff-varphi}\\
 P\left(\frac{\partial Q}{\partial\theta}\right)^2
   -Q\left(\frac{\partial P}{\partial\theta}\right)^2 &=& 0. \label{diff-theta}
\end{eqnarray}}

Let us look at the general case --- the quantum channel with parameters satisfying $a_0a_1a_2a_3a_4\sin\mu\neq 0$. Suppose $\sin\theta\neq 0$,
and $P(\theta,\varphi)Q(\theta,\varphi)\neq 0$.

By (\ref{diff-theta-varphi}), there is
\begin{equation}\label{general-diff-theta-varphi}
\begin{array}{ll}
  & 2b_1x^2+(b_2\cos\varphi+b_3\sin\varphi)x \\
& -a_0^2b_1+b_4\cos2\varphi+b_5\sin2\varphi =  0,
\end{array}
\end{equation}
where {\small
\begin{eqnarray*}
x &=& a_0\cot\theta, ~~  g_1= a_2a_3a_4,\\
 g_2 &=& a_2^2a_3^2-(a_2^2+a_3^2)a_4^2, ~~ g_3 =2a_1^2+2a_4^2 -1,\\
  b_1 & =& a_1g_1\sin\mu, ~~  b_2= 2a_1\sin\mu(3a_1g_1\cos\mu-g_2), \\
 b_3 &=& 2a_1g_2\cos\mu+g_1(1-2a_0^2-a_1^2-3a_1^2\cos2\mu-2a_4^2),  \\
b_4  &=& a_1\sin\mu[g_1(a_0^2-1+4a_1^2\cos^2\mu+2a_4^2)-2a_1g_2\cos\mu], \\
b_5  &=& g_1^2-a_1g_1[a_1^2\cos 3\mu-(a_2^2+a_3^2-a_4^2)\cos\mu]+a_1^2g_2\cos2\mu.
\end{eqnarray*} }
Eq.(\ref{diff-varphi}) subtracted from Eq.(\ref{diff-theta}) is {\small
\begin{equation}\label{general-diff-theta-diff-varphi}
\begin{array}{rl}
 8d_1x^3+4x^2(d_2\sin\varphi+d_3\cos\varphi)+2x(d_4+d_5\cos2\varphi & \\
 +d_6\sin2\varphi)+d_7\cos\varphi+d_8\sin\varphi+d_9\cos3\varphi+d_{10}\sin3\varphi & =  0,
\end{array}
\end{equation}
where
\begin{eqnarray*}
  d_1 &=& a_1g_1(g_3+a_0^2)\cos\mu-a_1^2g_2-g_1^2, \\
 d_2 &=& 2 a_1\sin\mu[a_1g_1(2g_3+a_0^2)\cos\mu -2g_1^2
 +(a_0^2-2a_1^2)g_2], \\
  d_3 &=& g_1(3a_0^2-1-2a_0^4+4a_1^2a_4^2+4a_2^2a_3^2-2a_0^2a_4^2\\
  & & +4 a_1^2g_3\cos^2\mu-2 a_0^2a_1^2\sin^2\mu)\\
&& -2 a_1\cos\mu[6g_1^2-(a_0^2-2a_1^2)g_2], \\
  d_4 &=& 2a_2^2a_3^2(a_4^2-2a_2^2a_3^2-3a_1^2a_4^2)+(1-2a_0^2-2a_1^2+4a_0^2a_1^2\\
&& -2a_1^4)g_2-16a_1^2g_1^2\cos^2\mu
  +2a_1g_1\cos\mu(5a_0^2-2-2a_0^4\\
&& +a_1^2-3a_0^2a_1^2+2a_1^4+8a_2^2a_3^2-2a_0^2a_4^2+6a_1^2a_4^2), \\
d_5 &=& 4(a_0^2-a_1^2)g_1^2+a_1^3g_1(g_3-a_0^2)\cos3\mu+a_1g_1\cos\mu[2a_0^2\\
&& -(1-2a_0^2)^2+a_1^2(g_3+4a_4^2-5a_0^2)+4a_2^2a_3^2-8a_0^2a_4^2]\\
~~&~~&+2 a_1^2\cos2\mu[(2a_0^2-a_1^2)g_2-3g_1^2],\\
d_6 &=& a_1\sin\mu \{g_1[2a_0^2(1+2a_2^2+2a_3^2-2a_4^2)-1+4a_2^2a_3^2 \\
&& +4 a_1^2a_4^2+4a_1^2(g_3-a_0^2)\cos^2\mu]+4a_1[(2a_0^2-a_1^2)g_2 \\
&& -3g_1^2]\cos\mu\}, \\
d_7 &=& g_1\{1-2a_0^4a_1^2-4a_2^2a_3^2-2a_4^2+a_1^2(4a_4^2-2) \\
&&  -a_0^2[2 +2a_1^4-4a_4^2-a_1^2(3-6a_4^2)]-2 a_1^2\cos^2\mu[2a_0^4 \\
&& +3-12a_2^2a_3^2-4a_1^2(1+a_4^2)+a_0^2(5a_1^2-7+2a_4^2)] \\
~~&~~&-4 a_1^3g_1\cos3\mu\}+2 a_1\cos\mu\{(1-2a_0^2-2a_1^2\\
&& +3a_0^2a_1^2)g_2-2a_2^2a_3^2[2a_2^2a_3^2-(1+a_0^2-5a_1^2)a_4^2]\},\\
d_8 &=&2 a_1\sin\mu[(1-2a_0^2-2a_1^2+3a_0^2a_1^2)g_2\\
&& -2a_2^2a_3^2(2a_2^2a_3^2-a_4^2+a_0^2a_4^2+4a_1^2a_4^2) \\
&& -4 a_1^2g_1^2\cos2\mu+a_1g_1\cos\mu(7a_0^2-3-2a_0^4 \\
&& +4a_1^2-5a_0^2a_1^2+12a_2^2a_3^2-2a_0^2a_4^2+4a_1^2a_4^2)], \\
d_9 &=& a_0^2a_1\{2a_1^2g_2\cos3\mu-g_1[ a_1\cos2\mu(2a_0^2-3+3a_1^2\\
&& +6a_4^2)+ a_1^3\cos4\mu-4g_1\cos\mu ]\}, \\
d_{10} &=&   a_0^2a_1\{2a_1^2g_2\sin3\mu-g_1[ a_1\sin2\mu(2a_0^2-3+3a_1^2\\
&& +6a_4^2)+ a_1^3\sin4\mu-4g_1\sin\mu ]\}.
 \end{eqnarray*} }
Note that  $b_1\neq 0$. By dividing the left side of (\ref{general-diff-theta-diff-varphi}) by the left side of
(\ref{general-diff-theta-varphi}), we obtain the polynomial remainder
\begin{equation}\label{x-eq.}
\begin{array}{c}
  -xk_1(\varphi)+k_2(\varphi) =0, \\
\end{array}
\end{equation}
where  {\small
\begin{eqnarray*}
k_1(\varphi) &=& c_7+c_5\cos2\varphi+c_6\sin2\varphi,\\
k_2(\varphi) &=& c_1\cos3\varphi +c_2\sin3\varphi+c_3\sin\varphi+c_4\cos\varphi,\\
  c_1 &=& -b_2b_4d_1+b_3b_5d_1-b_1b_5d_2+b_1b_4d_3-b_1^2d_9, \\
 c_2 &=& -b_3b_4d_1-b_2b_5d_1-b_1^2d_{10}+b_1b_4d_2+b_1b_5d_3 ,\\
 c_3 &=& b_3b_4d_1-b_2b_5d_1-b_1b_4d_2+b_1b_5d_3-b_1^2d_8\\
 && +2b_1b_3d_1a_0^2-2b_1^2d_2a_0^2,
 \end{eqnarray*}
\begin{eqnarray*}
 c_4 &=& -b_2b_4d_1-b_3b_5d_1+b_1b_5d_2+b_1b_4d_3-b_1^2d_7\\
 && +2b_1b_2d_1a_0^2-2b_1^2d_3a_0^2, \\
 c_5 &=& b_2^2d_1-b_3^2d_1-4b_1b_4d_1+b_1b_3d_2-b_1b_2d_3+2b_1^2d_5, \\
c_6 &=& 2b_2b_3d_1-4b_1b_5d_1-b_1b_2d_2-b_1b_3d_3+2b_1^2d_6, \\
c_7 &=& b_2^2d_1+b_3^2d_1-b_1b_3d_2-b_1b_2d_3+2b_1^2d_4+4b_1^2d_1a_0^2.
 \end{eqnarray*}   }
 Using some algebra, we can prove that any $\varphi$ such that $\sin\varphi=0$ is not
minimum point of $R(\theta,\varphi)$. Thus, we can suppose  $\sin\varphi\neq 0$.

We first discuss the case $k_1(\varphi)\neq 0$.   From (\ref{x-eq.}), there is $x=\frac {k_2(\varphi)}{k_1(\varphi)}$.  Substituting it in
Eq.(\ref{general-diff-theta-varphi}), we derive {\small
\begin{equation}\label{v-eq.}
 \begin{array}{l}
 V^6[2b_1(c_1+c_4)^2+b_2(c_5+c_7)(c_1+c_4)+(c_5+c_7)^2(b_4-b_1a_0^2)]\\
 +V^5\{(c_1+c_4)[4b_1(3c_2+c_3)+2b_2c_6+b_3(c_5+c_7)]\\
 +(c_5+c_7)[b_2(3c_2+c_3)+2b_5(c_5+c_7)+4c_6(b_4-b_1a_0^2)]\}\\
 -V^4\{2b_1[8c_1^2-(3c_2+c_3)^2-2(c_1-c_4)^2]-2b_3(c_1+c_4)c_6\\
 +2(b_2c_1-4b_5c_6)(c_5+c_7)+2b_2(c_1c_5-c_4c_7)\\
 -(c_5-c_7)^2
 (b_4+b_1a_0^2)-(3c_2+c_3)(b_3c_5+2b_2c_6+b_3c_7)\\
 +4b_4(c_5^2-c_6^2)+4b_1(c_6^2+c_7^2)a_0^2\}\\
 -2V^3[4b_1c_1(5c_2+c_3)-(c_2+c_3)(4b_1c_4+b_2c_7)\\
 +2(b_3c_1+b_2c_2)c_5+2b_5(c_5^2-c_7^2)+b_2(3c_1-c_4)c_6\\
 -b_3(3c_2+c_3)c_6+b_3(c_1-c_4)c_7+4c_6(b_4c_5-b_5c_6+b_1c_7a_0^2)]\\
 +V^2 \{2b_1[(3c_1-c_4)^2-8c_2^2+2(c_2+c_3)^2]-4b_3c_2c_5\\
 +(3c_1-c_4)[b_2(c_5-c_7)-2b_3c_6]+2c_7b_3(c_2+c_3)\\
 +2b_2(c_3-c_2)c_6+4b_4(c_5^2-c_6^2)-8b_5c_6(c_5-c_7)\\
 +(c_5+c_7)^2(b_1a_0^2-b_4)-4b_1(c_6^2+c_7^2)a_0^2\}\\
 +V [(c_2-c_3)(12b_1c_1-4b_1c_4+b_2c_5-2b_3c_6-b_2c_7)\\
 +(c_5-c_7)(3b_3c_1-b_3c_4+2b_5c_5+4b_4c_6-2b_5c_7+4b_1c_6a_0^2)]\\
 =0,
 \end{array}\end{equation}  }
where $V=\cot\varphi$.
   Finding out  the solution(s)
$(\theta_i,\varphi_i)$ of (\ref{v-eq.}) and (\ref{x-eq.}) satisfying the two equations in
(\ref{differencial-theta-and-varphi}) if there exists, then determining the minimum point $(\theta_1,
\varphi_1)$  such that $\min\{R(\theta_i,\varphi_i)\}=R(\theta_1,\varphi_1)$ if there is, we obtain
\begin{equation*}\label{general-min}
R_{\min}=\{
\begin{array}{lll}
  \min\{R(\theta_1,\varphi_1), Q(\theta_0, \varphi_0), R(0,\varphi) \}, & \texttt{if} & a_2=a_3, \\
 \min\{R(\theta_1,\varphi_1), R(0,\varphi)\},& \texttt{if} & a_2\neq a_3.
 \end{array}
\end{equation*}
Otherwise, there is no minimum point at the case  $P(\theta,\varphi)Q(\theta,\varphi)\sin\theta\neq 0$, and the
minimum
\begin{equation}\label{general-min-1}
 R_{\min} =\{
 \begin{array}{lll}
  \min\{ Q(\theta_0, \varphi_0), R(0,\varphi) \}, & \texttt{if} & a_2=a_3, \\
  R(0,\varphi),& \texttt{if} & a_2\neq a_3.
 \end{array}
\end{equation}

Now let us look at the case  $k_1(\varphi)=0$.   If it has  common solution(s) with  $k_2(\varphi)=0$, and there
is/are solutions/solutions $(\theta_j, \varphi_j)$ of Eqs.(\ref{general-diff-theta-varphi}) and
(\ref{general-diff-theta-diff-varphi}) satisfying Eqs.(\ref{differencial-theta-and-varphi}), then we have
\begin{equation*}\label{general-min-2}
 R_{\min}=\{
\begin{array}{lll}
 \min\{R(\theta_2, \varphi_2),  Q(\theta_0,\varphi_0), R(0,\varphi) \}, & \texttt{if} & a_2=a_3, \\
 \min\{R(\theta_2, \varphi_2),  R(0,\varphi) \}, & \texttt{if} & a_2\neq a_3,
\end{array}
\end{equation*}
where $R(\theta_2, \varphi_2)=\min\{R(\theta_j, \varphi_j)\}$. Otherwise, the minimum $R_{\min}$ is the same as
that in (\ref{general-min-1}).

Note that in  the three expressions of $R_{\min}$ above, we use the properties $R(0,\varphi)=R(\pi,\varphi)$ and
$Q(\theta_0,\varphi_0)=P(\theta_{0'},\varphi_{0'})$, where
$P(\theta_0,\varphi_0)=Q(\theta_{0'},\varphi_{0'})=0$.

For the quantum channel (\ref{quantumchannel}) with $a_1a_2a_3a_4\sin\mu=0$, we also obtain the exact values of the maximal successful
probabilities for
 controlled teleportation.

According to the definition  $E_{ij}=\max_\varepsilon \sum_s p_s E(|\phi_s\rangle)$   of LE in
\cite{VPCirac-prl(04)},  the maximal probability $p_{\max}$ in (\ref{max}) is a kind of LE. If
$E(|\phi_s\rangle)$   is chosen to be the concurrence  of $|\phi_s\rangle$,
 we  show  the exact value of the LE
\begin{eqnarray*}\label{maxLE}
E_{23}&=&\max\{p_1C_1+p_2C_2\} \\
 &=& 2\sqrt{a_2^2a_3^2-2a_1a_2a_3a_4\cos\mu+(a_0^2+a_1^2)a_4^2}.
\end{eqnarray*}

In conclusion, we have shown the sufficient and necessary condition that a three-qubit state can be collapsed to an EPR pair by an appropriate
measurement on one qubit. We also characterized the tripartite states that can be used for perfect controlled teleportation. Moreover, we gave
the maximal  successful probability for controlled teleportation via a general tripartite state, and determined the exact value of another
localizable entanglement in \cite{VPCirac-prl(04)}.

The authors thank Prof. J. I. Cirac for his fruitful discussions and for his hospitality during their  stay at Max-Planck-Institut f\"{u}r
Quantenoptik. This work was supported by the National Natural Science Foundation (NSF) of China under Grant No: 10671054, Hebei NSF of China
under Grant Nos: A2005000140, 07M006, and the Key Project of Science and Technology Research of Education Ministry of China under Grant No:
207011.

\end{document}